\documentclass[10pt,twocolumn,english,aps,final,pre]{revtex4}
\usepackage{mathptmx}
\usepackage[T1]{fontenc}
\usepackage[latin9]{inputenc}
\usepackage{babel}

\usepackage{amsmath}
\usepackage{graphicx}
\usepackage{esint}
\usepackage[unicode=true, pdfusetitle,
 bookmarks=true,bookmarksnumbered=false,bookmarksopen=false,
 breaklinks=false,pdfborder={0 0 1},backref=false,colorlinks=false]
 {hyperref}

\makeatletter

\newcommand{\lyxdot}{.}

\@ifundefined{textcolor}{}
{%
 \definecolor{BLACK}{gray}{0}
 \definecolor{WHITE}{gray}{1}
 \definecolor{RED}{rgb}{1,0,0}
 \definecolor{GREEN}{rgb}{0,1,0}
 \definecolor{BLUE}{rgb}{0,0,1}
 \definecolor{CYAN}{cmyk}{1,0,0,0}
 \definecolor{MAGENTA}{cmyk}{0,1,0,0}
 \definecolor{YELLOW}{cmyk}{0,0,1,0}
 }

\makeatother

\begin{document}

\title{Propagation Of Waves In Periodic-Heterogeneous Bistable Systems}

\author{Jakob Löber$^{1},$ Markus Bär$^{2},$ and Harald Engel$^{1}$}

\affiliation{$^{1}$Institut für Theoretische Physik, Technische Universität Berlin,
Hardenbergstrasse 36, 10623 Berlin, Germany}

\affiliation{$^{2}$Physikalisch-Technische Bundesanstalt, Department 8.4 Mathematical
Modelling and Data Analysis, Abbestr. 2-12, 10587 Berlin, Germany}
\begin{abstract}
Wave propagation in one-dimensional heterogeneous bistable media is
studied using the Schlögl model as a representative example. Starting
from the analytically known traveling wave solution for the homogeneous
medium, infinitely extended, spatially periodic variations in kinetic
parameters as the excitation threshold, for example, are taken into
account perturbatively. Two different multiple scale perturbation
methods are applied to derive a differential equation for the position
of the front under perturbations. This equation allows the computation
of a time independent average velocity, depending on the spatial period
length and the amplitude of the heterogeneities. The projection method
reveals to be applicable in the range of intermediate and large period
lengths but fails when the spatial period becomes smaller than the
front width. Then, a second order averaging method must be applied.
These analytical results are capable to predict propagation failure,
velocity overshoot, and the asymptotic value for the front velocity
in the limit of large period lengths in qualitative, often quantitative
agreement with the results of numerical simulations of the underlying
reaction-diffusion equation. Very good agreement between numerical
and analytical results has been obtained for waves propagating through
a medium with periodically varied excitation threshold.
\end{abstract}
\maketitle

\section{Introduction}

\subsection{Heterogeneities in Reaction-Diffusion Systems}

Spatio-temporal patterns of Reaction-Diffusion Systems (RDS) are of
fundamental interest in many chemical \cite{kapral1995chemical} and
biological systems \cite{keener2008mathematical}. Prominent examples
are the Belousov-Zhabotinsky reaction (BZR), action potential propagation
in cardiac tissue and chemical catalysis. A large variety of patterns
can be found, e.g. traveling fronts and pulses in one spatial dimension,
spiral waves and traveling spots in two and three spatial dimensions
\cite{cross1993pfo,bode2002interaction}. Most of the models describing
these effects are assumed spatially homogeneous, although at least
biological systems are intrinsically heterogeneous. Recently there
is an increasing interest in heterogeneous RDS where the diffusion
coefficients, the reaction rate constants or other important parameters
as the excitation threshold depend explicitly on space. Effects of
heterogeneities on traveling wave solutions of RDS reach from reflection
and diffraction to velocity overshoots (the velocity of the wave is
larger in the case of heterogeneities than without them) and propagation
failure or oscillatory pinning \cite{schutz1995bfd,teramoto2009onset,keener2000pwe,bode1997fbr,xin2000fph,nishiura2007dtppppp}.
In two spatial dimensions, breakup of plane waves into spiral waves
and spatio-temporal chaos can occur \cite{bub2002spiral,bar2002pfa}.\\
The BZ reaction is an example for an experimental realization of
a heterogeneous RDS. The use of masks and lithographic techniques
permits the introduction in the reaction of patterned illumination
and patterned distribution of catalyst, respectively. A spatially
varied intensity of applied light corresponds to a spatial variation
of the excitation threshold of the system. A mathematical model for
this reaction is the modified Oregonator model \cite{krug1990amc}.
The velocity of pulse propagation in one spatial dimension under the
influence of a spatially periodically rectangularly varied excitation
threshold was studied numerically in \cite{schebesch1998wph} and
revealed a velocity overshoot at small period lengths. An analytical
investigation by Keener \cite{keener2000pwe} based on the averaging
theorem for the Schlögl model with a spatial variation of a reaction
rate in form of a Dirac comb showed a large velocity overshoot for
period lengths of the heterogeneities smaller than the frontwidth.
We derive a slightly generalized version of Keener's method in section
III.\\
An alternative perturbation method was applied to scalar and multicomponent
RDS by Bode \cite{bode1997fbr,schutz1995bfd} and also by Nishiura
\cite{nishiura2007dtppppp,teramoto2009onset} with mostly bump-type
heterogeneities. In section II a different form of this perturbation
method, called projection method throughout this publication, is proposed,
which allows the investigation of the effect of heterogeneities with
medium and large period lengths on front propagation. In the form
we use this method it was also applied in e.g. \cite{engel1985nif,mikhailov1983smp,schimanskygeier1983efp}
to traveling fronts in stochastic bistable media. Both methods are
applied to various variations of kinetic parameters of the Schlögl
model in section IV, and compared with numerical results in section
V.

\subsection{The Schlögl model}

This scalar RDS, proposed by Zeldovich-Frank-Kamenetsky \cite{zel1938k}
as a model for front propagation, and later applied by Schlögl \cite{schlogl1972crm}
as a model for a non-equilibrium phase transition, also known as bistable
equation, is given in its general form as \begin{equation}
\partial_{t}u=D\partial_{x}^{2}u+R\left(u\right)\label{eq:RDSGeneral}\end{equation}
with a reaction function\begin{align}
R\left(u\right) & =-k\left(u-u_{1}\right)\left(u-u_{2}\right)\left(u-u_{3}\right),\,0\leq u_{1}<u_{2}<u_{3}.\label{eq:SchloeglGeneral}\end{align}
The parameters $u_{1},\, u_{3}$ are stable fixed points and the parameter
$u_{2}$ is an unstable fixed point which corresponds to the excitation
threshold of the system and $D$ is the diffusion coefficient. $k$
is called reaction coefficient and has a dimension $\left[\textrm{time}\right]^{-1}.$
It is a measure of the intrinsic time scale on which the reaction
takes place. The traveling front solution of \eqref{eq:SchloeglGeneral}
is \cite{mikhailov1990fs}\begin{multline}
u\left(x,t\right)=U_{c}\left(\xi\right)=\\
\frac{1}{2}\left(u_{1}+u_{3}+\left(u_{1}-u_{3}\right)\tanh\left(\frac{1}{2}\sqrt{\frac{k}{2D}}\left(u_{3}-u_{1}\right)\xi\right)\right)\label{eq:FrontProfileGeneralSchloeglModel}\end{multline}
 with $\xi=x-ct$ and a velocity\begin{align}
c & =\sqrt{\frac{Dk}{2}}\left(u_{1}+u_{3}-2u_{2}\right).\label{eq:VelocityGeneralSchloeglModel}\end{align}
The front width $l$ of the traveling wave solution can be defined
as \cite{schlogl1972crm}\begin{align}
l & =\frac{4\sqrt{2D}}{\sqrt{k}\left(u_{3}-u_{1}\right)}.\end{align}
For every choice of the value of the excitation threshold $u_{2}$
the front has a certain velocity $c=c\left(u_{2}\right)$ but the
\textit{same} front profile $U_{c},$ which shows no explicit dependence
on $u_{2}.$ This is a peculiarity of the Schlögl model.The general
Schlögl model \eqref{eq:SchloeglGeneral} can be cast, without loss
of generality, into the form\begin{align}
\partial_{t}u & =\partial_{x}^{2}u-u\left(u-u_{2}\right)\left(u-1\right),\;0<u_{2}<1.\label{eq:SchloeglStandard}\end{align}
The traveling wave solution, with boundary conditions\begin{align}
\lim_{x\rightarrow\infty} & u\left(x,t\right)=0, & \lim_{x\rightarrow-\infty}u\left(x,t\right)=1,\end{align}
simplifies to \begin{align}
U_{c}\left(\xi\right) & =\frac{1}{2}\left(1-\tanh\left(\frac{1}{2\sqrt{2}}\xi\right)\right)=\frac{1}{1+e^{\frac{\xi}{\sqrt{2}}}}\end{align}
with a velocity\begin{align}
c & =\dfrac{1}{\sqrt{2}}\left(1-2u_{2}\right).\end{align}
All analytical computations are done with the simpler form \eqref{eq:SchloeglStandard}
of the Schlögl model.

\subsection{Harmonic mean velocity}

The appropriate average speed for a wave traveling with a space dependent
velocity is the harmonic mean of the velocity. Suppose a wave travels
a distance $L/2$ with velocity $c_{1}$ and the same distance $L/2$
with velocity $c_{2}.$ The total time it takes the wave to travel
through both regions is \begin{align}
T & =T_{1}+T_{2}=\frac{L}{2}\left(\frac{1}{c_{1}}+\frac{1}{c_{2}}\right)\end{align}
and the harmonic mean velocity is \begin{align}
\bar{c}_{harm} & =\frac{L}{T}=\frac{2}{1/c_{1}+1/c_{2}}.\label{eq:zwodifferentvelocities}\end{align}
If the velocity is assumed to depend on the space coordinate $x$
in a periodic way with period length $1,$ $c=c\left(x\right),$ one
can approximate the average velocity over an arbitrary period length
$L$ by dividing $L$ in $n$ pieces \begin{align}
\bar{c}_{harm} & =\frac{n}{\sum_{i=1}^{n}1/c\left(x_{i}/L\right)}.\end{align}
Consider a traveling wave with speed $c$ depending on a parameter
$s$, $c=c\left(s\right)$. If this parameter is spatially varied
in the form of an infinitely extended periodic function with period
$L,$ $s=s\left(x/L\right),$ the harmonic mean velocity can be approximated
as \begin{align}
\bar{c}_{harm} & =\frac{n}{\sum_{i=1}^{n}1/c\left(s\left(x_{i}/L\right)\right)}.\end{align}
The underlying assumption is that the front instantaneously adapts
its velocity when transiting from one region of space to the other,
distinguished by the different values $s\left(x_{i}/L\right)$ of
the parameter $s.$ In the limit of infinitely many pieces the sum
becomes an integral and the harmonic mean velocity is \begin{align}
\bar{c}_{harm} & =1/\left(\intop_{0}^{1}\frac{1}{c\left(s\left(x\right)\right)}\textrm{d}x\right).\label{eq:HarmonicMeanParameterS}\end{align}
For the more realistic case that it takes the front some transient
time to adapt its velocity when transiting from one region to the
other, one can state that $\bar{c}_{harm}$ \eqref{eq:HarmonicMeanParameterS}
still gives a good approximation for the average speed, if the transient
time is small compared to the traveling time through one period of
the spatial heterogeneities. This is always the case for the limit
of an infinite period length of the spatial heterogeneities. Thus
one expects the harmonic mean velocity \eqref{eq:HarmonicMeanParameterS}
to give the approximate average velocity for a wave traveling through
a periodic medium with large period lengths.

\section{Projection Method}

Consider an unperturbed scalar RDS in one spatial dimension, Eq. \eqref{eq:RDSGeneral}
(with $D=1$) and a traveling wave solution $U_{c}\left(\xi\right)$
with constant velocity $c.$ A perturbation $\kappa\left(u,x,t\right),$
depending on space and time as well as on $u,$ is introduced. $\kappa$
is multiplied by $\epsilon,$ which serves as the small parameter
for the perturbation expansion and is set to $\epsilon=1$ at the
end of the computations. The perturbed RDS, written in the co-moving
frame $\xi=x-ct$ of the unperturbed RDS \eqref{eq:RDSGeneral} and
with $u=u\left(\xi,t\right)$, is \begin{align}
\partial_{t}u & =\partial_{\xi}^{2}u+c\partial_{\xi}u+R\left(u\right)+\epsilon\kappa\left(u,\xi+ct,t\right).\label{eq:PerturbedRDS}\end{align}
The introduction of an additional heterogeneous diffusion coefficient
is straightforward \cite{kulka1995iir}, but not done here. Under
the perturbation $\kappa,$ the approximate solution of the perturbed
RDS \eqref{eq:PerturbedRDS} is assumed to be \begin{align}
u\left(\xi,t\right) & =U_{c}\left(\xi\right)+\epsilon\tilde{u}\left(\xi,t\right).\label{eq:UcPlusCorrection}\end{align}
Inserting \eqref{eq:UcPlusCorrection} into the unperturbed RDS \eqref{eq:RDSGeneral}
and expanding $R$ in powers of $\epsilon$ leads, in order $\epsilon^{1},$
to \begin{align}
\partial_{t}\tilde{u} & =\partial_{\xi}^{2}\tilde{u}+c\partial_{\xi}\tilde{u}+R'\left(U_{c}\left(\xi\right)\right)\tilde{u}=\mathcal{L}\tilde{u}.\end{align}
The operator $\mathcal{L}$ is a linear differential operator with
eigenvalues $\lambda$ which determine the stability of the traveling
wave solution $U_{c}.$ Under the condition of a translationally invariant
reaction function in \eqref{eq:RDSGeneral} one can prove the existence
of a certain eigenfunction $\tilde{u}\left(\xi,t\right)=U_{c}'\left(\xi\right)$
of $\mathcal{L}$ corresponding to the eigenvalue $\lambda_{0}=0$
which is called the Goldstone mode. If the wave is stable, this is
the largest eigenvalue. The function $\tilde{u}$ can be split up
in a part parallel to the Goldstone mode and and a part orthogonal
to it (with $p,\, q$ arbitrary constants), \begin{align}
\tilde{u}\left(\xi,t\right) & =pU_{c}'\left(\xi\right)+qv\left(\xi,t\right)\label{eq:PerturbationSplit}\end{align}
with \begin{align}
\left\langle W^{\dagger}\left(\xi\right),v\left(\xi,t\right)\right\rangle  & =\intop_{-\infty}^{\infty}W^{\dagger}\left(\xi\right)v\left(\xi,t\right)\textrm{d}\xi=0.\label{eq:ProjectionCondition}\end{align}
Eq. \eqref{eq:ProjectionCondition} is called projection condition,
where \begin{align}
\left\langle g,f\right\rangle  & =\intop_{-\infty}^{\infty}g\left(\xi\right)f\left(\xi\right)\textrm{d}\xi\label{eq:InnerProduct}\end{align}
is the inner product in function space and $W^{\dagger}$ is the Goldstone
mode of the adjoint operator of $\mathcal{L},$ $\mathcal{L}^{\dagger}W^{\dagger}=0,$
and will be referred to as the adjoint Goldstone mode. The Goldstone
mode $U_{c}'$ leads to a small shift of the traveling wave solution,\begin{align}
U_{c}\left(\xi\right)+pU_{c}'\left(\xi\right)+qv\left(\xi,t\right) & \approx U_{c}\left(\xi+p\right)+qv\left(\xi,t\right),\end{align}
so the projection condition \eqref{eq:ProjectionCondition} can be
interpreted as saying that $v$ does not take part in a shift of the
traveling wave solution, it only leads to a deformation of the wave
profile.\\
A slow time scale $T=\epsilon t$ is introduced and the time derivative
transformed accordingly\begin{align}
\partial_{t} & \rightarrow\partial_{t}+\partial_{t}T\partial_{T}=\partial_{t}+\epsilon\partial_{T}.\label{eq:TransformationTimeDerivative}\end{align}
From now on, $T$ and $t$ are treated as independent of each other,
as it is the usual procedure in multiple scale perturbation theory.
The ansatz for the solution of the perturbed problem \eqref{eq:PerturbedRDS}
is\begin{align}
u\left(\xi,t,T\right) & =U_{c}\left(\xi+p\left(T\right)\right)+\epsilon v\left(\xi,t,T\right),\end{align}
together with the projection condition\begin{multline}
\left\langle W^{\dagger}\left(\xi+p\left(T\right)\right),v\left(\xi,t,T\right)\right\rangle =\\
\intop_{-\infty}^{\infty}W^{\dagger}\left(\xi+p\left(T\right)\right)v\left(\xi,t,T\right)\textrm{d}\xi=0,\label{eq:ProjectionConditionRDS}\end{multline}
where the correction to the position of the front under perturbation
$p\left(T\right)$ is constant on the original time scale $t$ but
depending on the 2nd timescale $T.$ The functions $R\left(u\right)$
and $\kappa\left(u,\xi+ct,t\right)$ are expanded in powers of $\epsilon.$
In order $\epsilon^{1}$ one gets\begin{multline}
-\partial_{t}v\left(\xi,t,T\right)+\mathcal{L}v\left(\xi,t,T\right)=\\
-\kappa\left(U_{c}\left(\xi+p\left(T\right)\right),\xi+ct,t\right)+p'\left(T\right)U_{c}'\left(\xi+p\left(T\right)\right),\label{eq:FrontShapeDiffEq}\end{multline}
a linear PDE for the correction $v$ of the front shape under perturbation
with an inhomogeneity on the r.h.s. Eq. \eqref{eq:FrontShapeDiffEq}
is projected onto the adjoint Goldstone mode, and by applying a variant
of the projection condition \eqref{eq:ProjectionConditionRDS} one
can eliminate one term to get\begin{multline}
\left\langle W^{\dagger}\left(\xi+p\left(T\right)\right),\mathcal{L}v\left(\xi,t,T\right)\right\rangle =\\
-\left<W^{\dagger}\left(\xi+p\left(T\right)\right),\kappa\left(U_{c}\left(\xi+p\left(T\right)\right),\xi+ct,t\right)\right.\\
\left.\vphantom{W^{\dagger}\left(\xi+p\left(T\right)\right),\kappa\left(U_{c}\left(\xi+p\left(T\right)\right),\xi+ct,t\right)}-p'\left(T\right)U_{c}'\left(\xi+p\left(T\right)\right)\right>\label{eq:DiffEqProjected}\end{multline}
This turns out to be a solvability condition (or Fredholm alternative,
see \cite{keener2000pam}) for $v,$ which guarantees a bounded solution
for $v$ only if the r.h.s. of \eqref{eq:DiffEqProjected} is zero.
With the adjoint Goldstone mode of a scalar traveling wave solution
$U_{c}\left(\xi\right)$ in one spatial dimension,\begin{align}
W^{\dagger}\left(\xi\right) & =e^{c\xi}U_{c}'\left(\xi\right),\end{align}
and a coordinate change $\xi\rightarrow\xi+p\left(T\right),$ the
solvability condition reads as\begin{multline}
\intop_{-\infty}^{\infty}e^{c\xi}U_{c}'\left(\xi\right)\left(\kappa\left(U_{c}\left(\xi\right),\xi+ct-p\left(T\right),t\right)\right.\\
\left.\vphantom{\kappa\left(U_{c}\left(\xi\right),\xi+ct-p\left(T\right),t\right)}-p'\left(T\right)U_{c}'\left(\xi\right)\right)\textrm{d}\xi=0.\end{multline}
This is a differential equation for $p\left(T\right),$ the correction
to the position of the front. Rescaling $p\left(T\right)$ to the
original time $t$ by introducing a new function \begin{align}
\phi\left(t\right) & =ct-p\left(T\right)\label{eq:PhiDef}\end{align}
and using \eqref{eq:TransformationTimeDerivative} gives\begin{align}
\frac{\textrm{d}}{\textrm{d}t}\phi\left(t\right) & =c-\epsilon\frac{\textrm{d}}{\textrm{d}T}p\left(T\right).\end{align}
Thus one has derived the ODE for the position of the traveling wave
under perturbation\begin{align}
\frac{\textrm{d}}{\textrm{d}t}\phi\left(t\right) & =c+\epsilon\Theta_{1}^{P}\left(\phi\left(t\right)\right)\nonumber \\
\Theta_{1}^{P}\left(\phi\left(t\right)\right) & =\frac{1}{K_{c}}\intop_{-\infty}^{\infty}e^{c\xi}U_{c}'\left(\xi\right)\kappa\left(U_{c}\left(\xi\right),\xi+\phi\left(t\right),t\right)\textrm{d}\xi,\label{eq:ODEPositionOfFrontProjectionMethod}\end{align}
with \begin{align}
K_{c} & =\intop_{-\infty}^{\infty}e^{c\xi}U_{c}'^{2}\left(\xi\right)\textrm{d}\xi.\label{eq:Kc}\end{align}
 The advantage of introducing a new function $\phi,$ \eqref{eq:PhiDef},
is that if the perturbation $\kappa$ has no explicit time dependence,
one gets an autonomous ODE for $\phi\left(t\right)$ which is usually
much easier to solve. If the r.h.s. of \eqref{eq:ODEPositionOfFrontProjectionMethod}
is zero, propagation fails. \\
In the case of a time independent perturbation $\kappa$ with period
length $L$ \begin{align}
\kappa\left(u,x\right) & =\kappa\left(u,x+L\right),\end{align}
one can compute a time-independent average velocity $c_{avg}$ over
one period of the heterogeneities. The traveling time $T_{c}$ is
the time it takes the front to travel through one period of the spatially
periodic heterogeneities\begin{align}
T_{c} & =\intop_{0}^{L}\frac{\textrm{d}\phi}{c+\epsilon\Theta_{1}^{P}\left(\phi\right)}=L\intop_{0}^{1}\frac{\textrm{d}\phi}{c+\epsilon\Theta_{1}^{P}\left(\phi L\right)}\end{align}
and the average velocity is\begin{align}
c_{avg} & =\frac{L}{T_{c}}=1/\intop_{0}^{1}\frac{\textrm{d}\phi}{c+\epsilon\Theta_{1}^{P}\left(\phi L\right)}.\label{eq:AverageVelocity}\end{align}

\section{Averaging Method}

Keener developed a method based on the averaging theorem \cite{bogoliubov1961asymptotic,keener2000pam}
to derive an ODE for the position of the front of a RDS with a heterogeneous
reaction function of the form \begin{align}
R\left(u,x\right) & =\left(1+g'\left(\frac{x}{L}\right)\right)f\left(u\right)-a\left(u\right),\label{eq:HeterogeneousReactionFunction}\end{align}
where $f$ is an arbitrary nonlinear function \cite{keener2000pwe}.
He also treated the case of a heterogeneous diffusion coefficient
\cite{keener2000hap} with a similar method. The spatial heterogeneities
have to be periodic and the period length $L$ is used as the small
parameter for the perturbation expansion.\\
Written as a system of two differential equations with reaction
function \eqref{eq:HeterogeneousReactionFunction}, \eqref{eq:RDSGeneral}
is\begin{align}
\partial_{x}u & =v,\nonumber \\
\partial_{x}v & =\partial_{t}u-\left(1+g'\left(\frac{x}{L}\right)\right)f\left(u\right)+a\left(u\right).\label{eq:HeterogeneousRDSTwoDiffEq}\end{align}
In contrast to Keener's approach \cite{keener2000pwe}, which assumes
a linear dependence of $a$ on $u,$ here the function $a\left(u\right)$
is allowed to be an arbitrary, possibly nonlinear function. The function
$g'\left(x\right)$ denotes the heterogeneities with period one and
zero mean:\begin{equation}
\left\langle g'\left(x\right)\right\rangle =\intop_{0}^{1}g'\left(x\right)\textrm{d}x=0.\end{equation}
A second length scale $\tau=x/L$ is introduced and the space derivative
transformed accordingly, $\partial_{x}\rightarrow\partial_{x}+\frac{1}{L}\partial_{\tau}.$
An exact change of variables from $u,\, v$ to new variables $y,\, z$
is applied to \eqref{eq:HeterogeneousRDSTwoDiffEq} to get, after
expanding nonlinear terms in \eqref{eq:HeterogeneousRDSTwoDiffEq}
and the unknown functions $y,\, z$ in powers of $L,$ a homogeneous
averaged system\begin{equation}
\partial_{t}y_{0}\left(x,t\right)=\partial_{x}^{2}y_{0}\left(x,t\right)+f\left(y_{0}\left(x,t\right)\right)-a\left(y_{0}\left(x,t\right)\right)\label{eq:HomoEq}\end{equation}
in order $L^{0}$ and linear inhomogeneous PDEs for the new variables
in higher orders of $L.$ Eq. \eqref{eq:HomoEq} must have an analytically
known traveling wave solution $y_{0}\left(x,t\right)=U_{c}\left(x-ct\right).$
Transforming into the co-moving frame $\xi=x-\phi\left(t\right)$
with an unknown, time dependent position of the front $\phi\left(t\right),$
and applying a solvability condition to the linear PDE obtained in
the next non-vanishing order of $L$ yields an ODE for $\phi\left(t\right).$
For more details, see Keener's publication \cite{keener2000pwe}.

\subsection{Averaging in 1st order}

The exact change of variables is\begin{align}
u\left(x,t,\tau\right) & =y\left(x,t,\tau\right)\nonumber \\
v\left(x,t,\tau\right) & =z\left(x,t,\tau\right)-Lg\left(\tau\right)f\left(u\right),\label{eq:ExactChangeOfVariables1stOrderAveraging}\end{align}
where $g$ is the anti derivative of $g'.$ The ODE for the position
of the front under perturbation is \begin{align}
\frac{\textrm{d}\phi\left(t\right)}{\textrm{d}t} & =c+L\Theta_{1}^{A}\left(\phi\left(t\right)\right)\label{eq:DiffEqAveraging1stOrder}\end{align}
with \begin{equation}
\Theta_{1}^{A}\left(\phi\right)=\frac{1}{K_{c}}\intop_{-\infty}^{\infty}g\left(\frac{\xi+\phi}{L}\right)\frac{\textrm{d}}{\textrm{d}\xi}\left(f\left(U_{c}\right)U_{c}'e^{c\xi}\right)\textrm{d}\xi\label{eq:Theta1}\end{equation}
and $K_{c}$ given as above \eqref{eq:Kc}.

\subsection{Averaging in 2nd order}

For second order averaging a different exact change of coordinates
is applied which does not only eliminate all heterogeneous terms in
order $L^{0},$ but also all terms in order $L^{1}$ and yields a
linear inhomogeneous PDE for the new variables $y_{2},\, z_{2}$ in
order $L^{2}.$\\
We follow Keener and use as exact change of coordinates \begin{align}
u\left(x,t,\tau\right) & =y\left(x,t\right)-L^{2}G\left(\tau\right)f\left(y\right),\nonumber \\
v\left(x,t,\tau\right) & =z\left(x,t\right)-Lg\left(\tau\right)f\left(y\right)+L^{2}G\left(\tau\right)f'\left(y\right)y_{x}\nonumber \\
 & +L^{3}g\left(\tau\right)G\left(\tau\right)f'\left(y\right)f\left(y\right),\label{eq:ExactChangeOfVariables2ndOrderAveraging}\end{align}
where $G$ is the anti derivative of $g.$ If \begin{align}
\lim_{\xi\rightarrow-\infty}f\left(U_{c}\left(\xi\right)\right) & =0,\label{eq:ConditionIntegrationConstant}\end{align}
the integration constants for $g$ and $G$ can be chosen so that
the mean values $\left\langle g\left(x\right)\right\rangle $ and
$\left\langle G\left(x\right)\right\rangle $ are zero. If not, one
has to choose\begin{align}
\left\langle G\left(x\right)\right\rangle  & =\left\langle g^{2}\left(x\right)\right\rangle \lim_{\xi\rightarrow-\infty}\dfrac{f'\left(U_{c}\right)}{f'\left(U_{c}\right)-a'\left(U_{c}\right)}.\label{eq:IntegrationConstant}\end{align}
 The ODE for the position of the front under perturbation in 2nd order
averaging is\begin{equation}
\frac{\textrm{d}\phi\left(t\right)}{\textrm{d}t}=c+L^{2}\Theta_{2}^{A}\left(\phi\right),\label{eq:DiffEqAveraging2ndOrder}\end{equation}
with \begin{align}
\Theta_{2}^{A}\left(\phi\right) & =-\dfrac{1}{K_{c}}\intop_{-\infty}^{\infty}g^{2}\left(\frac{\xi+\phi}{L}\right)f'\left(U_{c}\right)f\left(U_{c}\right)U_{c}'e^{c\xi}\textrm{d}\xi\nonumber \\
 & -\intop_{-\infty}^{\infty}G\left(\frac{\xi+\phi}{L}\right)\frac{\textrm{d}}{\textrm{d}\xi}\left(e^{c\xi}\frac{\textrm{d}}{\textrm{d}\xi}\left(f\left(U_{c}\right)U_{c}'\right)\right)\textrm{d}\xi.\label{eq:Theta2}\end{align}
The explicit dependence on the homogeneous part $a\left(u\right)$
of the reaction function can be eliminated in both cases \eqref{eq:DiffEqAveraging1stOrder},
\eqref{eq:DiffEqAveraging2ndOrder}, which is also the reason why
these results are the same as derived by Keener \cite{keener2000pwe}
for a linear function $a\left(u\right).$ Eq. \eqref{eq:Theta1} and
\eqref{eq:Theta2} depend implicitly on $a$ through the traveling
wave solution of the homogeneous case $U_{c}\left(\xi\right).$\\
To compute an average velocity $c_{avg}$ over one period length
of the heterogeneities, one proceeds analogously to the projection
method \eqref{eq:AverageVelocity}, and derives\begin{align}
c_{avg} & =\frac{L}{T_{c}}=1/\intop_{0}^{1}\frac{\textrm{d}\phi}{c+L^{2}\Theta_{2}^{A}\left(\phi L\right)}.\label{eq:AverageVelocity2ndOrderAveraging}\end{align}

\subsection{Equivalence of 1st order averaging and projection method}

With one partial integration and assuming that $g\left(\frac{\xi+\phi}{L}\right)f\left(U_{c}\right)U_{c}'e^{c\xi}$
vanishes at the boundaries, the ODE for the position of the front
derived in first order averaging \eqref{eq:DiffEqAveraging1stOrder}
becomes

\begin{align}
\frac{\textrm{d}\phi\left(t\right)}{\textrm{d}t} & =c-\frac{1}{K_{c}}\intop_{-\infty}^{\infty}g'\left(\frac{\xi+\phi}{L}\right)f\left(U_{c}\right)U_{c}'e^{c\xi}\textrm{d}\xi\label{eq:ODEPosFrontAve1stOrder}\end{align}
If the periodic-heterogeneous reaction function can be split up in
the form we assumed for the averaging method, $R\left(u,x\right)=f\left(u\right)-a\left(u\right)+g'\left(x/L\right)f\left(u\right)=R\left(u\right)+\kappa\left(u,x\right),$
then the ODE for the position of the front derived with the projection
method \eqref{eq:ODEPositionOfFrontProjectionMethod} becomes

\begin{align}
\frac{\textrm{d}}{\textrm{d}t}\phi\left(t\right) & =c-\frac{\epsilon}{K_{c}}\intop_{-\infty}^{\infty}g'\left(\frac{\xi+\phi}{L}\right)f\left(U_{c}\right)U_{c}'e^{c\xi}\textrm{d}\xi.\label{eq:ODEPosFrontProjectionMethodPeriodicHetterogeneities}\end{align}
For evaluation of \eqref{eq:ODEPosFrontProjectionMethodPeriodicHetterogeneities},
$\epsilon$ is set to $\epsilon=1$ and one realizes that \eqref{eq:ODEPosFrontProjectionMethodPeriodicHetterogeneities}
and \eqref{eq:ODEPosFrontAve1stOrder} are equal. Note the differences
in the approaches of these two multiple scale perturbation expansions:
for the projection method, a second time scale $T=\epsilon t$ is
introduced. For the averaging method a second space scale $\tau=x/L$
is introduced and the heterogeneities are restricted to periodic ones
with small period lengths. Thus it can be shown that the validity
of equation \eqref{eq:DiffEqAveraging1stOrder} for the position of
the front derived in 1st order averaging can be extended to arbitrary
large period lengths and even to non-periodic heterogeneities. In
fact, in the case of the Schlögl model, it fails for small period
lengths but gives good results for large period lengths, as will be
shown later.

\section{Analytical Results for the Schlögl Model}

\subsection{Projection method for a variation of $u_{1},\, u_{2},\, u_{3}$ and
$k$}

An infinitely extended sinusoidal variation of the excitation threshold
$u_{2}$ of the Schlögl model is considered, \begin{align}
u_{2}\left(x\right) & =u_{2}+A\sin\left(2\pi x/L\right),\end{align}
where $A$ is the amplitude of the spatial variation and $L$ its
period length. The heterogeneous reaction function is\begin{align}
R\left(u,x\right) & =-u\left(u-\left(u_{2}+A\sin\left(2\pi x/L\right)\right)\right)\left(u-1\right)\nonumber \\
 & =-u\left(u-u_{2}\right)\left(u-1\right)+A\sin\left(2\pi x/L\right)u\left(u-1\right)\nonumber \\
 & =R\left(u\right)+\kappa\left(u,x\right).\label{eq:HetReacFuncU2}\end{align}
Similarly, one introduces the heterogeneities for the variations of
$u_{1},\, u_{3}$ and $k.$ The derivation of the ODE for the position
of the front under perturbation can be done simultaneously for all
four variations by introducing a general perturbation\begin{align}
\kappa\left(u,x\right) & =\left(u-Z_{1}\right)\left(u-Z_{2}\right)\left(Z_{4}u+Z_{3}\right)A\sin\left(2\pi x/L\right),\end{align}
where one has for the different variations
\begin{enumerate}
\item $u_{1}\left(x\right)=A\sin\left(2\pi x/L\right):$

$Z_{1}=1,\, Z_{2}=u_{2},\, Z_{3}=1,\, Z_{4}=0,$

\item $u_{2}\left(x\right)=u_{2}+A\sin\left(2\pi x/L\right):$

$Z_{1}=1,\, Z_{2}=0,\, Z_{3}=1,\, Z_{4}=0,$

\item $u_{3}\left(x\right)=1+A\sin\left(2\pi x/L\right):$

$Z_{1}=0,\, Z_{2}=u_{2},\, Z_{3}=1,\, Z_{4}=0,$

\item $k\left(x\right)=1+A\sin\left(2\pi x/L\right):$

$Z_{1}=1,\, Z_{2}=u_{2},\, Z_{3}=0,\, Z_{4}=-1.$

\end{enumerate}
Note that the amplitude $A$ is restricted to values for which the
condition $0\leq u_{1}<u_{2}<u_{3}$ and $k>0$ is fulfilled locally,
e.g. for a variation of $u_{2}:$ $u_{1}<u_{2}\left(x\right)<u_{3}$
for all $x.$ This implies that the computation for a variation of
$u_{1}$ cannot be done with the form \eqref{eq:SchloeglStandard}
of the Schlögl model, where $u_{1}=0,$ but the general form \eqref{eq:SchloeglGeneral}
has to be used instead.\\
The differential equation for the position of the front $\phi\left(t\right),$
see \eqref{eq:ODEPositionOfFrontProjectionMethod}, is \begin{align}
\frac{\textrm{d}}{\textrm{d}t}\phi\left(t\right) & =c-\frac{\epsilon A}{K_{c}}\intop_{-\infty}^{\infty}e^{c\xi}U_{c}'\left(\xi\right)\sin\left(2\pi\left(\xi+\phi\left(t\right)\right)/L\right)\nonumber \\
 & \times\left(U_{c}\left(\xi\right)-Z_{1}\right)\left(U_{c}\left(\xi\right)-Z_{2}\right)\left(Z_{4}U_{c}\left(\xi\right)+Z_{3}\right)\textrm{d}\xi\nonumber \\
 & =c+\epsilon C_{1}\left(C_{2}\sin\left(\frac{2\pi\phi(t)}{L}\right)+C_{3}\cos\left(\frac{2\pi\phi(t)}{L}\right)\right).\label{eq:ODEPositionOfFrontSchloeglModel}\end{align}
The values of the constants $C_{1},\, C_{2}$ and $C_{3}$ are given
in the appendix, see \eqref{eq:Cs}. The average velocity $c_{avg}$
can be computed according to the formula \eqref{eq:AverageVelocity}
to get\begin{align}
c_{avg} & =\sqrt{c^{2}-\epsilon^{2}C_{1}^{2}\left(C_{2}^{2}+C_{3}^{2}\right)}.\label{eq:CavgGeneral}\end{align}
The ODE for the position of the front \eqref{eq:ODEPositionOfFrontSchloeglModel}
with the initial condition $\phi\left(0\right)=0$ can be solved to
give\begin{align}
\phi\left(t\right) & =\frac{L}{\pi}\arctan\left(\frac{c+\epsilon C_{1}C_{3}}{\cot\left(\frac{\pi}{L}c_{avg}t\right)c_{avg}-\epsilon C_{1}C_{2}}\right),\end{align}
which describes the position of the front for one spatial period of
the heterogeneities, $-\frac{L}{2}<\phi\left(t\right)<\frac{L}{2}.$\\
Propagation failure occurs if $\dfrac{\textrm{d}}{\textrm{d}t}\phi\left(t\right)=0.$
With the help of the r.h.s. of \eqref{eq:ODEPositionOfFrontSchloeglModel}
one derives the relation %
\footnote{determine the value of $\phi$ for which the r.h.s. of \eqref{eq:ODEPositionOfFrontSchloeglModel}
attains its minimum, substitute this value back into the r.h.s. of
\eqref{eq:ODEPositionOfFrontSchloeglModel} and determine a relation
between the parameters under which the r.h.s. of \eqref{eq:ODEPositionOfFrontSchloeglModel}
is zero.%
}\begin{align}
c_{avg} & =0\label{eq:ConditionPropagationFailure}\end{align}
as the condition for propagation failure.\\
Computation of the ODE for the position of the front \eqref{eq:ODEPositionOfFrontProjectionMethod}
can be done for periodic arbitrary shaped heterogeneities by expanding
$\kappa\left(u,x\right)$ in a Fourier series in $x.$ This was done
for variations of all kinetic parameters in form of an infinitely
extended rectangular function, but results are not shown.

\subsection{Failure of projection method for small period lengths }

As was already mentioned by Keener \cite{keener2000hap,keener2000pwe},
1st order averaging fails for small period lengths and smooth nonlinear
functions $f\left(u\right).$ This can be shown for the Schlögl model
by estimating the dependence of the coefficients $C_{1}C_{2}$ and
$C_{1}C_{3}$ in \eqref{eq:ODEPositionOfFrontSchloeglModel} on $L,$
see \eqref{eq:Cs}. For small $L,$ one can approximate \begin{align}
\cosh\left(\frac{1}{L}\right) & \approx\frac{1}{2}e^{1/L},\\
\sinh\left(\frac{1}{L}\right) & \approx\frac{1}{2}e^{1/L}\end{align}
in $C_{2},\, C_{3},$ and in the denominator of $C_{1}$\begin{align}
\left|\cos\left(2\sqrt{2}c\pi\right)-\cosh\left(\frac{4\sqrt{2}\pi^{2}}{L}\right)\right| & \leq\left|1+\cosh\left(\frac{4\sqrt{2}\pi^{2}}{L}\right)\right|\\
 & \approx\frac{1}{2}e^{\frac{4\sqrt{2}\pi^{2}}{L}}.\nonumber \end{align}
By keeping only the largest terms up to order $L^{1}$ in $H_{1},\, H_{2}$
(\eqref{eq:HsProj1}, \eqref{eq:HsProj2}), one derives\begin{multline}
c^{2}-c_{avg}^{2}=\epsilon^{2}C_{1}^{2}\left(C_{2}^{2}+C_{3}^{2}\right)\\
=\frac{256A^{2}\epsilon^{2}e^{-\frac{4\sqrt{2}\pi^{2}}{L}}\pi^{6}\sin^{2}\left(\sqrt{2}c\pi\right)}{\left(c-2c^{3}\right)^{2}L^{8}}\left(4\pi^{2}\sin^{2}\left(\sqrt{2}c\pi\right)Z_{4}^{2}\right.\\
\left.\vphantom{4\pi^{2}\sin^{2}\left(\sqrt{2}c\pi\right)Z_{4}^{2}}+\left(2\pi\cos\left(\sqrt{2}c\pi\right)Z_{4}+L\sin\left(\sqrt{2}c\pi\right)\right.\right.\\
\left.\left.\times\left(\left(4c+\sqrt{2}\left(2Z_{1}+2Z_{2}-3\right)\right)Z_{4}-2\sqrt{2}Z_{3}\right)\right)^{2}\right)+O\left(L^{2}\right)\label{eq:Approx}\end{multline}
The term $e^{-4\sqrt{2}\pi^{2}/L}$ vanishes faster than any polynomial
order of $L,$ and is called transcendentally small by Keener.

\subsection{2nd order averaging for a sinusoidally varied excitation\\threshold}

The heterogeneous reaction function is cast in the form necessary
for averaging \eqref{eq:HeterogeneousReactionFunction} \begin{align}
R\left(u,x\right) & =-u\left(u-\left(u_{2}+A\sin\left(2\pi x/L\right)\right)\right)\left(u-1\right)\nonumber \\
 & =\left(1+\dfrac{A}{u_{2}}\sin\left(2\pi x/L\right)\right)u\left(u-1\right)u_{2}-u^{2}\left(u-1\right)\nonumber \\
 & =\left(1+g'\left(\dfrac{x}{L}\right)\right)f\left(u\right)-a\left(u\right).\end{align}
To derive the differential equation \eqref{eq:DiffEqAveraging2ndOrder}
one has to solve the integrals arising in $\Theta_{2}^{A},$ \eqref{eq:Theta2}.
The integration constants for the anti derivative $g$ of $g'$ and
$G$ of $g$ can be chosen so that $g$ and $G$ have vanishing mean,
\eqref{eq:ConditionIntegrationConstant}. The function $\Theta_{2}^{A}\left(\phi\right)$
can be computed analytically to get\begin{multline}
\Theta_{2}^{A}\left(\phi\right)=H_{1}\sin\left(\frac{2\pi\phi}{L}\right)+H_{2}\cos\left(\frac{2\pi\phi}{L}\right)\\
+H_{3}\sin\left(\frac{4\pi\phi}{L}\right)+H_{4}\cos\left(\frac{4\pi\phi}{L}\right)+H_{5}.\label{eq:Theta2VarU2}\end{multline}
The values of the constants $H_{1}$ up to $H_{5}$ can be found in
the appendix, see \eqref{eq:Hsu2}. The average velocity \eqref{eq:AverageVelocity2ndOrderAveraging}
is computed numerically, because analytical computation is possible,
but very tedious.\\
Computation of the ODE for the position of the front \eqref{eq:ODEPositionOfFrontProjectionMethod}
can be done for periodic arbitrary shaped variations of the excitation
threshold $u_{2}$ by expanding $g$ and $G$ in \eqref{eq:Theta2}
in a Fourier series in $x.$ This was done for a rectangular variation
of the excitation threshold, but results are not shown.%
\begin{figure}[t]
\begin{center}

\textit{\includegraphics[width=0.9\columnwidth]{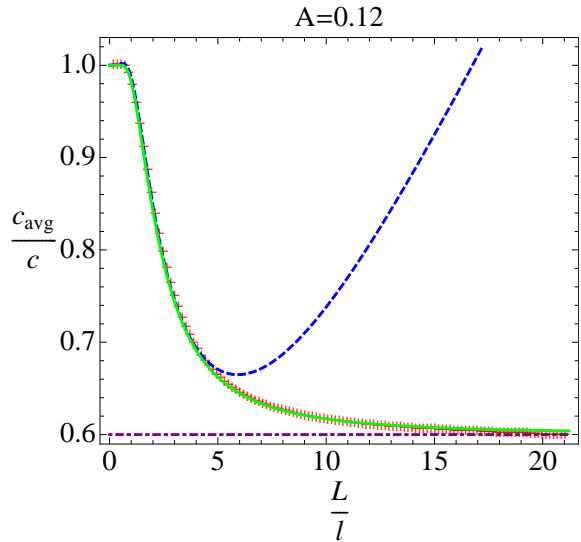}}

\end{center}\caption[{\small Sinusoidal variation of $u_{2}$ for large period lengths}]{\textit{}{\small Ratio of velocities $c_{avg}/c$ plotted over the
ratio of period length $L$ and frontwidth $l$ for a sinusoidal variation
of excitation threshold $u_{2}\left(x\right)=0.35+A\sin\left(2\pi x/L\right).$
Projection method (green solid line) shows excellent agreement with
numerical results (red crosses), 2nd order averaging (blue dashed
line) fails for large period lengths. The harmonic mean velocity (purple
dash-dotted line) gives the approximate average velocity for large
period lengths.}\label{fig:CavgU2SinusLarge}}

\end{figure}

\subsection{2nd order averaging for a sinusoidally varied fixed point parameter
$u_{3}$}

The heterogeneous reaction function \eqref{eq:HeterogeneousReactionFunction}
is \begin{align}
R\left(u,x\right) & =-u\left(u-u_{2}\right)\left(u-\left(1+A\sin\left(2\pi x/L\right)\right)\right)\nonumber \\
 & =\left(1+A\sin\left(2\pi x/L\right)\right)u\left(u-u_{2}\right)-u^{2}\left(u-u_{2}\right)\nonumber \\
 & =\left(1+g'\left(\dfrac{x}{L}\right)\right)f\left(u\right)-a\left(u\right).\end{align}
The integration constant for $G$ is determined so that $\left\langle G\right\rangle =\frac{A^{2}\left(u_{2}-2\right)}{8\pi^{2}\left(3u_{2}-5\right)},$
\eqref{eq:IntegrationConstant}. The ODE for the position of the front
has the same form as for a variation of $u_{2},$

\begin{multline}
\Theta_{2}^{A}\left(\phi\right)=H_{1}\sin\left(\frac{2\pi\phi}{L}\right)+H_{2}\cos\left(\frac{2\pi\phi}{L}\right)\\
+H_{3}\sin\left(\frac{4\pi\phi}{L}\right)+H_{4}\cos\left(\frac{4\pi\phi}{L}\right)+H_{5},\end{multline}
The constants $H_{1}$ up to $H_{5}$ are listed in the appendix,
see \eqref{eq:Hsu3}. The average velocity \eqref{eq:AverageVelocity2ndOrderAveraging}
is computed numerically.%
\begin{figure}[t]
\begin{center}

\textit{\includegraphics[width=1\columnwidth]{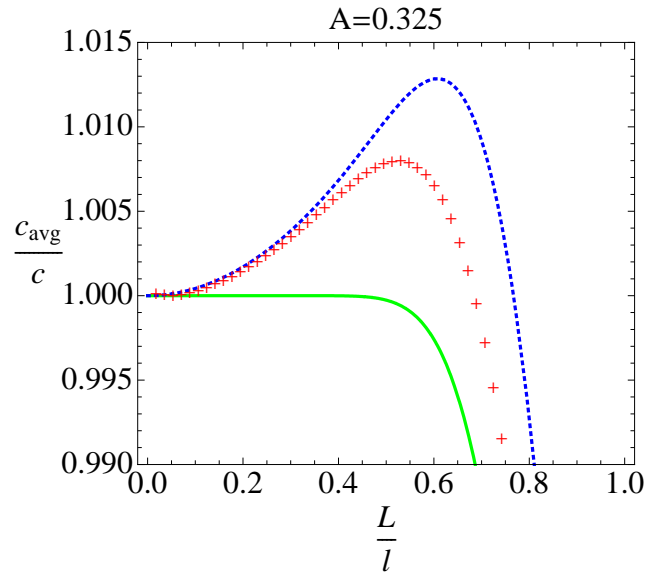}}

\end{center}\caption[{\small Sinusoidal variation of $u_{2}$ for small period lengths}]{{\small Velocity overshoot for a sinusoidal variation of excitation
threshold $u_{2}\left(x\right)=0.35+A\sin\left(2\pi x/L\right),$
predicted by 2nd order averaging (blue dashed line) in qualitative
agreement with numerical results (red crosses). Projection method
(green solid line) fails for small period lengths.}\textit{\label{fig:CavgU2SinusSmall}}}

\end{figure}
\begin{figure}
\begin{center}

\includegraphics[width=0.94\columnwidth]{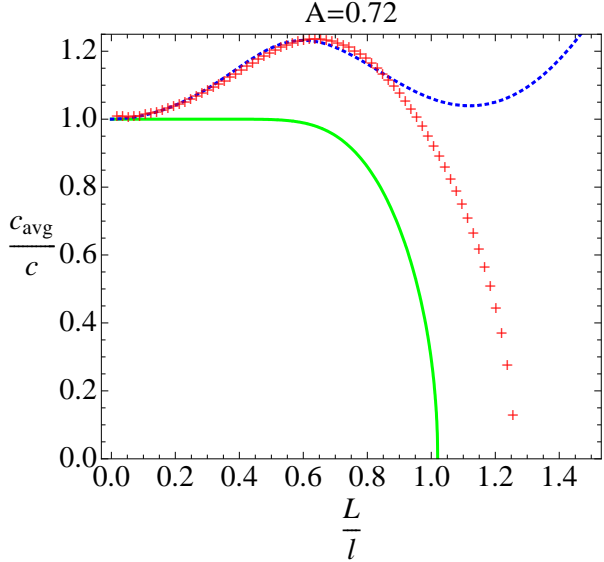}

\end{center}

\caption[{\small Sinusoidal variation of $u_{3}$}]{{\small Velocity overshoot} {\small for a sinusoidal variation of
$u_{3}\left(x\right)=1+A\sin\left(2\pi x/L\right)$ and $u_{2}=0.35$
predicted by 2nd order averaging (blue dashed line) in agreement with
numerical results (red crosses). Propagation failure is predicted
qualitatively by projection method (green solid line), which fails
for small period lengths.}\label{fig:CavgU3Sinus}}

\end{figure}

\section{Comparison of Analytical and Numerical Results}

The PDE for the heterogeneous Schlögl model is solved numerically
with a simple finite differences Euler forward scheme and the average
velocity is obtained by a linear fit of the space over time data over
an integer multiple of period lengths. Data near to the boundaries
have to be neglected because of boundary effects.

\subsection{Variation of excitation threshold $u_{2}$}

A sinusoidally varied excitation threshold $u_{2}\left(x\right)=u_{2}+A\sin\left(\frac{2\pi x}{L}\right)$
leads to an average velocity\begin{multline}
c_{avg,\, u_{2}}=\left(c^{2}-\frac{4A^{2}\left(\left(u_{2}-1\right)^{2}L^{2}+2\pi^{2}\right)}{\left(2u_{2}^{3}-3u_{2}^{2}+u_{2}\right)^{2}L^{6}}\right.\\
\left.\vphantom{c^{2}-\frac{4A^{2}\left(\left(u_{2}-1\right)^{2}L^{2}+2\pi^{2}\right)}{\left(2u_{2}^{3}-3u_{2}^{2}+u_{2}\right)^{2}L^{6}}}\times\frac{\left(u_{2}^{2}L^{2}+2\pi^{2}\right)\left(L^{2}\left(1-2u_{2}\right)^{2}+8\pi^{2}\right)\sin^{2}\left(2u_{2}\pi\right)}{\left(\cosh\left(\frac{4\sqrt{2}\pi^{2}}{L}\right)-\cos\left(4u_{2}\pi\right)\right)}\right)^{\frac{1}{2}}\label{eq:CavgU2}\end{multline}
obtained by the projection method according to the formula \eqref{eq:CavgGeneral}.
For intermediate and large period lengths, the agreement between \eqref{eq:CavgU2}
and numerical results is excellent, see Fig. \ref{fig:CavgU2SinusLarge}.
The averaging method in 2nd order gives good results for small period
lengths but fails for large period lengths, as could be expected because
the period length is used as the small parameter for the perturbation
expansion (although 1st order averaging does not fail for large period
lengths).

\subsection{Velocity overshoot}

For period lengths smaller than the front width, the numerical results
for the sinusoidally varied excitation threshold show a small velocity
overshoot. The results obtained in 2nd order averaging predict this
overshoot qualitatively at the right size of period lengths. Eq. \eqref{eq:CavgU2}
shows a plateau-like behavior indicating the transcendental smallness
of the results obtained with the projection method for small period
lengths, see Fig. \ref{fig:CavgU2SinusSmall}.

For a sinusoidal variation of the fixed point parameter $u_{3}\left(x\right)=1+A\sin\left(\frac{2\pi x}{L}\right),$
a large velocity overshoot is found numerically, again predicted qualitatively
by 2nd order averaging and missed by the projection method, which
gives an analytical solution for the average velocity

\begin{multline}
c_{avg,\, u_{3}}=\left(c^{2}-\frac{A^{2}\left(\left(u_{2}-1\right)^{2}L^{2}+8\pi^{2}\right)}{\left(2u_{2}^{3}-3u_{2}^{2}+u_{2}\right)^{2}L^{6}}\right.\\
\left.\vphantom{c^{2}-\frac{A^{2}\left(u_{2}^{2}L^{2}+2\pi^{2}\right)\left(\left(u_{2}-1\right)^{2}L^{2}+8\pi^{2}\right)}{\left(2u_{2}^{3}-3u_{2}^{2}+u_{2}\right)^{2}L^{6}\left(\cosh\left(\frac{4\sqrt{2}\pi^{2}}{L}\right)-\cos\left(4u_{2}\pi\right)\right)}}\times\frac{\left(L^{2}\left(1-2u_{2}\right)^{2}+8\pi^{2}\right)\left(u_{2}^{2}L^{2}+2\pi^{2}\right)\sin^{2}\left(2u_{2}\pi\right)}{\left(\cosh\left(\frac{4\sqrt{2}\pi^{2}}{L}\right)-\cos\left(4u_{2}\pi\right)\right)}\right)^{\frac{1}{2}}.\label{eq:CavgU3}\end{multline}
\begin{figure}[t]
\begin{center}

\textit{\includegraphics[clip,width=0.9\columnwidth]{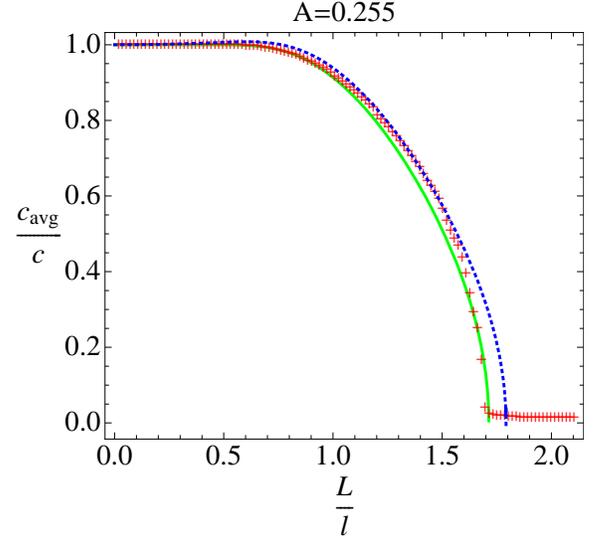}}

\end{center}

\caption[{\small Propagation failure for a sinusoidal variation of $u_{2}$}]{\textit{}{\small Propagation failure for a sinusoidal variation of
excitation threshold $u_{2}\left(x\right)=0.35+A\sin\left(2\pi x/L\right)$
occurs if the average velocity obtained by numerical simulation (red
crosses) drops to zero and can be predicted by projection method (green
solid line) and qualitatively by 2nd order averaging (blue dashed
line). \label{fig:PropFailureSinusU2Fig1}}}

\end{figure}
\begin{figure}[t]
\begin{center}\medskip{}

\textit{\includegraphics[clip,width=0.9\columnwidth]{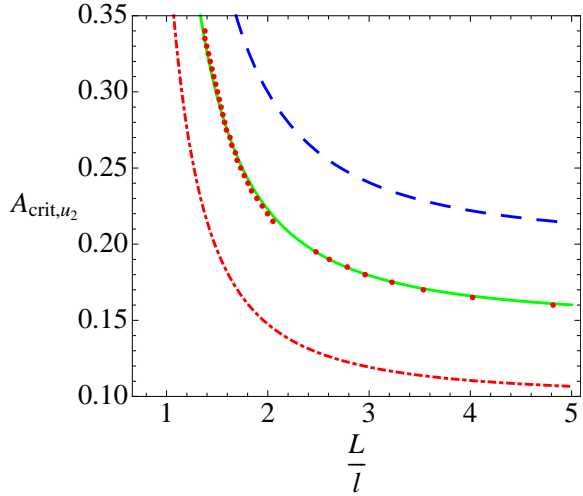}}

\end{center}

\caption[{\small Critical amplitude for which propagation failure occurs for
a sinusoidal variation of $u_{2}$}]{\textit{}{\small The critical amplitude for which propagation failure
occurs for a sinusoidal variation of the excitation threshold $u_{2}\left(x\right)=u_{2}+A\sin\left(2\pi x/L\right),$
obtained by the projection method, is a monotonically decreasing function
for all values of $u_{2}$ and shown for $u_{2}=0.3$ (blue dashed
line) and $u_{2}=0.4$ (red dash-dotted line). The results for a value
of $u_{2}=0.35$ (green solid line) are compared to numerical simulations
(red dots).}\textit{\small{} \label{fig:PropFailureSinusU2Fig2}}}

\end{figure}

\subsection{Propagation failure}

With the condition for propagation failure, $c_{avg}=0,$ the results
for the average velocity obtained with the projection method allows
to determine the critical amplitude for which propagation failure
occurs. For the case of a sinusoidally varied excitation threshold,
this gives \begin{multline}
A_{crit,\, u_{2}}\left(L\right)=\frac{L^{3}\sqrt{\cosh\left(\frac{4\sqrt{2}\pi^{2}}{L}\right)-\cos\left(4\pi u_{2}\right)}}{2\sqrt{2}\sqrt{L^{2}\left(u_{2}-1\right)^{2}+2\pi^{2}}}\\
\times\frac{\csc\left(2\pi u_{2}\right)\left(1-2u_{2}\right)^{2}\left(1-u_{2}\right)u_{2}}{\sqrt{L^{2}u_{2}^{2}+2\pi^{2}}\sqrt{L^{2}\left(1-2u_{2}\right)^{2}+8\pi^{2}}},\label{eq:U2PropFailureAL-2}\end{multline}
which is a monotonically decreasing function for all values of the
system parameter $u_{2}.$ This means that for a fixed amplitude large
enough, propagation failure occurs for all period lengths larger than
a certain critical period length, see Fig. \ref{fig:PropFailureSinusU2Fig1}.
For the choice of parameters shown in Fig. \ref{fig:PropFailureSinusU2Fig1},
the result of 2nd order averaging can predict the propagation failure,
but if propagation failure occurs at larger period lengths, it fails
badly to do so. Comparison of \eqref{eq:U2PropFailureAL-2} with numerical
results shows very good agreement, see Fig. \ref{fig:PropFailureSinusU2Fig2}.

\begin{figure}
\begin{center}

\includegraphics[width=0.9\columnwidth]{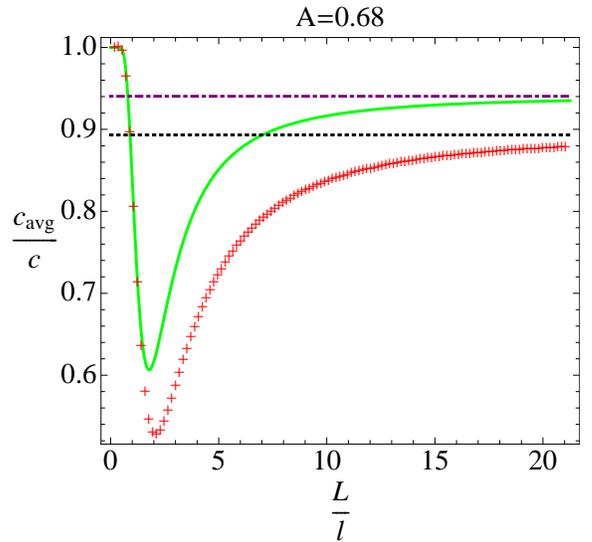}

\end{center}\caption[{\small Sinusoidal variation of $k$ for small amplitude}]{{\small Ratio of velocities $c_{avg}/c$ plotted over the ratio of
period length $L$ and front width $l$ for a sinusoidal variation
of $k\left(x\right)=1+A\sin\left(2\pi x/L\right)$ and $u_{2}=0.38.$
The average velocity shows a minimum and the numerical solution (red
crosses) approaches the harmonic mean velocity (black dashed line)
from below. The solution obtained by projection method (green solid
line) approaches a different limit (purple dot-dashed line).\label{fig:CavgKSinusFig1}}}

\end{figure}

\begin{figure}
\begin{center}

\includegraphics[width=0.9\columnwidth]{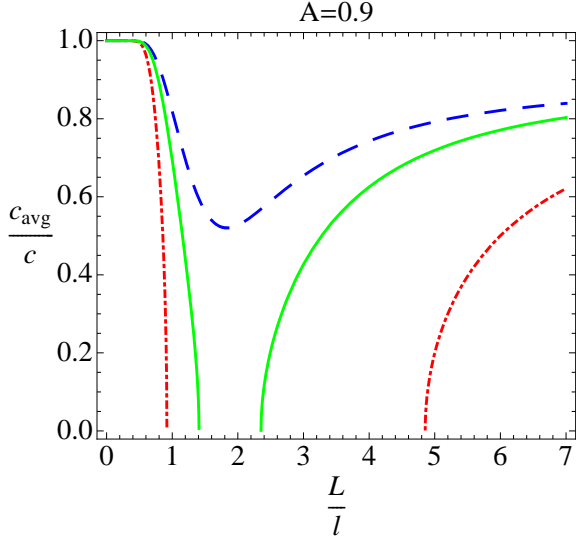}

\end{center}

\caption[{\small Sinusoidal variation of $k$ for different values of $u_{2}$}]{{\small The ratio of velocities $c_{avg}/c$ either shows an interval
of period lengths for which propagation failure occurs or a minimum
for a sinusoidal variation of $k\left(x\right)=1+A\sin\left(2\pi x/L\right),$
depending on the value of $u_{2}=0.35$ (blue dashed line), $u_{2}=0.425$
(red dot-dashed line), $u_{2}=0.38$ (green solid line).\label{fig:PropFailureSinusKFig1}}}

\end{figure}

\begin{figure}
\begin{center}\medskip{}

\includegraphics[width=0.9\columnwidth]{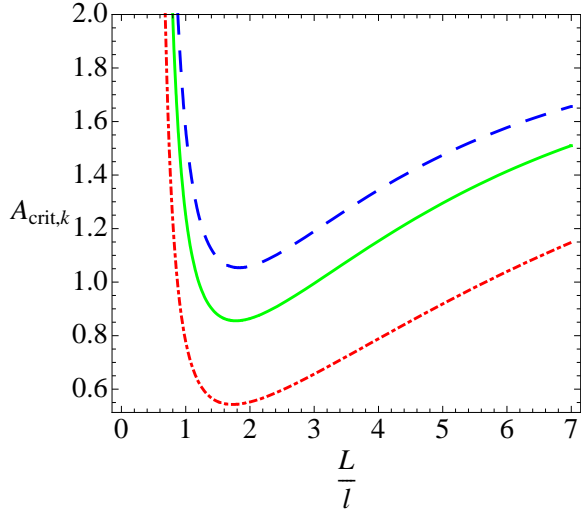}

\end{center}

\caption[{\small Critical amplitude for a sinusoidal variation of $k$}]{{\small The critical amplitude for which propagation failure occurs
obtained by projection method for a sinusoidal variation of $k\left(x\right)=1+A\sin\left(2\pi x/L\right)$
shows a minimum for all values of $u_{2}=0.35$ (blue dashed line),
$u_{2}=0.425$ (red dot-dashed line) and $u_{2}=0.38$ (green solid
line).\label{fig:PropFailureSinusKFig2}}}

\end{figure}

For a sinusoidal variation of the reaction coefficient $k\left(x\right)=1+A\sin\left(\frac{2\pi x}{L}\right)$
a different behavior occurs. The average velocity obtained with the
projection method is given as\begin{multline}
c_{avg,\, k}=\left(c^{2}-\frac{A^{2}\left(\left(u_{2}-1\right)^{2}L^{2}+2\pi^{2}\right)}{4u_{2}^{2}\left(u_{2}\left(2u_{2}-3\right)+1\right)^{2}}\right.\\
\left.\vphantom{c^{2}-\frac{A^{2}\left(\left(u_{2}-1\right)^{2}L^{2}+2\pi^{2}\right)}{4u_{2}^{2}\left(u_{2}\left(2u_{2}-3\right)+1\right)^{2}}}\times\frac{\left(u_{2}^{2}L^{2}+2\pi^{2}\right)\left(L^{2}\left(1-2u_{2}\right)^{2}+8\pi^{2}\right)^{2}\sin^{2}\left(2u_{2}\pi\right)}{L^{8}\left(\cosh\left(\frac{4\sqrt{2}\pi^{2}}{L}\right)-\cos\left(4u_{2}\pi\right)\right)}\right)^{\frac{1}{2}}\label{eq:CavgK}\end{multline}
and shows, for a certain range of the system parameter $u_{2}$ and
in qualitative agreement with numerical simulations, a minimum for
a finite period length, see Fig. \ref{fig:CavgKSinusFig1}. For a
value of the amplitude large enough and slightly different values
of the system parameter $u_{2},$ the projection method predicts propagation
failure occurring for an interval of period lengths, see Fig. \ref{fig:PropFailureSinusKFig1},
and propagation is possible for larger and smaller period lengths.
Comparison with numerical results shows qualitative agreement, see
Fig. \ref{fig:CavgKSinusFig2}. The critical amplitude for which propagation
failure occurs is given as \begin{multline}
A_{crit,\, k}\left(L\right)=\frac{\sqrt{2}L^{4}\sqrt{\cosh\left(\frac{4\sqrt{2}\pi^{2}}{L}\right)-\cos\left(4\pi u_{2}\right)}}{\sqrt{L^{2}\left(u_{2}-1\right)^{2}+2\pi^{2}}}\\
\times\frac{\csc\left(2\pi u_{2}\right)\left(1-2u_{2}\right)^{2}\left(1-u_{2}\right)u_{2}}{\sqrt{L^{2}u_{2}^{2}+2\pi^{2}}\left(L^{2}\left(1-2u_{2}\right)^{2}+8\pi^{2}\right)}.\label{eq:KPropFailureAL}\end{multline}
and is shown for different values of $u_{2}$ in Fig. \ref{fig:PropFailureSinusKFig2}.
There one can see again the fact that propagation fails for an interval
of period lengths: the critical amplitude has a minimum for a finite
period length and is not a monotonically decreasing function.

For most cases of a variation of $u_{3},$ the critical amplitude
for which propagation failure occurs is a monotonically decreasing
function, but for a very small range of the system parameter $u_{2},$
propagation failure can occur for an interval of period lengths (not
shown).%

\subsection{Front velocity in the limit of large period lengths}

The average velocity obtained with the projection method allows to
determine the limit for large period lengths. For the case of the
sinusoidal variation of excitation threshold $u_{2},$ this limit
is given as \begin{align}
\lim_{L\rightarrow\infty}c_{avg,\, u_{2}} & =\sqrt{c^{2}-2A^{2}},\label{eq:LimitLargePeriodCavgU2}\end{align}
which agrees with the harmonic mean of the velocities computed according
to \eqref{eq:HarmonicMeanParameterS}. The numerical solution approaches
the limit \eqref{eq:LimitLargePeriodCavgU2} from above, as shown
in Fig. \ref{fig:CavgU2SinusLarge}. The limit of the average velocity\begin{align}
\lim_{L\rightarrow\infty}c_{avg,\, u_{3}} & =\sqrt{c^{2}-\frac{1}{2}A^{2}},\end{align}
for the sinusoidal variation of $u_{3}$ also agrees with the harmonic
mean of the velocities and with numerical results (not shown).

\begin{figure}
\begin{center}

\includegraphics[width=0.9\columnwidth]{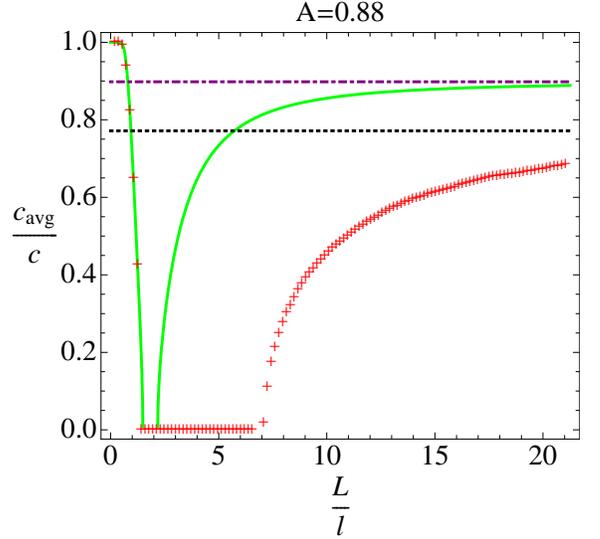}

\end{center}\caption[{\small Sinusoidal variation of $k$ for large amplitude}]{{\small Interval of period lengths for which propagation failure
occurs for a sinusoidal variation of reaction coefficient $k\left(x\right)=1+A\sin\left(2\pi x/L\right),$
predicted by projection method (green solid line) in qualitative agreement
with numerical results (red crosses). The limit for large period lengths
of the numerical results is given by the harmonic mean of the velocities
(black dashed line), the analytical solution approaches a different
limit (purple dot-dashed line).\label{fig:CavgKSinusFig2}}}

\end{figure}

For the sinusoidal variation of reaction coefficient $k$ the limit
for large period lengths of \eqref{eq:CavgK} is\begin{align}
\lim_{L\rightarrow\infty}c_{avg,\, k} & =c\sqrt{1-\frac{A^{2}}{4}},\label{eq:LimitLargePeriodCavgK}\end{align}
which does not agree with the harmonic mean of the velocities,\begin{align}
\bar{c}_{harm,\, k} & =\pi c\frac{1}{\left(\frac{K\left(\frac{2A}{A-1}\right)}{\sqrt{1-A}}+\frac{K\left(\frac{2A}{A+1}\right)}{\sqrt{1+A}}\right)},\label{eq:HarmMeanVarKappa}\end{align}
where $K\left(x\right)$ is the complete elliptic integral of the
first kind. The numerical solution approaches the harmonic mean of
the velocities \eqref{eq:HarmMeanVarKappa} from below, see Fig. \ref{fig:CavgKSinusFig1},
Fig. \ref{fig:CavgKSinusFig2}. The velocity of the unperturbed general
Schlögl model has the same square root dependence on the reaction
coefficient $k$ as on the diffusion coefficient $D,$ see \eqref{eq:VelocityGeneralSchloeglModel}.
It follows that the limit of large period lengths of the average velocity
for the case of a sinusoidal variation of the diffusion coefficient
$D\left(x\right)=1+A\sin\left(2\pi x/L\right)$ is the same as \eqref{eq:HarmMeanVarKappa},
which was checked by numerical simulations (not shown).

\section{Discussion}

Infinitely extended spatially varied reaction parameters of the Schlögl
model are considered and the effects on the propagation velocity are
studied. The applied perturbation methods seem to work best for a
variation of excitation threshold $u_{2},$ the reason is probably
that the front profile \eqref{eq:FrontProfileGeneralSchloeglModel}
does not depend on this parameter. The projection method works worse
for a variation of the reaction coefficient $k$ than for all other
variations, even not predicting the correct limit of the average velocity
for large period lengths. This could be connected to the fact that
the velocity of the homogeneous case \eqref{eq:VelocityGeneralSchloeglModel}
shows a linear dependence on the fixed point parameters but a square
root dependence on $k.$ The ODE for the position of the front obtained
in 1st order averaging is equivalent to the one obtained with the
projection method, and both fail generally for small period lengths
due to the transcendentally small dependence of $\Theta_{1}\left(\phi\right)$
on the period length. This causes the plateau in the plots of the
average velocity for small period lengths in all solutions obtained
with the projection method. The solutions obtained in 2nd order averaging
agree qualitatively with the numerical simulations and can predict
the size and the value of the period lengths for which velocity overshoots
occur, but generally fail for large period lengths. A small velocity
overshoot up to $1.5\%$ is found for a variation of the excitation
threshold $u_{2}$ with period lengths slightly smaller than the front
width. A similar velocity overshoot was found for period lengths for
a variation of the excitation threshold in the modified Oregonator
model \cite{schebesch1998wph}. For a variation of $u_{3},$ a larger
velocity overshoot up to $25\%$ is found, which occurs at period
lengths of approximately the same size as the velocity overshoot in
the case of the variation of $u_{2}.$ Propagation failure occurs
in both cases for amplitudes large enough and all period lengths larger
than a certain critical period length. For a variation of the reaction
coefficient $k,$ and for a small range of values of the system parameter
$u_{2}$ in the case of a variation of $u_{3},$ we find an interval
of period lengths for which propagation failure occurs. All computations
were done for a sinusoidal as well as for a rectangular variation
of the parameters, which show qualitatively the same effects. The
amplitude $A$ and the period length $L$ are more important in affecting
the front velocity than the shape of the heterogeneities.

\section*{Appendix A}

The ODE for the position of the front for a sinusoidal variation of
$u_{1},\, u_{2},\, u_{3}$ and $k$ obtained with the projection method
is

\begin{align}
\frac{\textrm{d}}{\textrm{d}t}\phi\left(t\right) & =c-\frac{\epsilon A}{K_{c}}\intop_{-\infty}^{\infty}e^{c\xi}U_{c}'\left(\xi\right)\sin\left(2\pi\left(\xi+\phi\left(t\right)\right)/L\right)\nonumber \\
 & \times\left(U_{c}\left(\xi\right)-Z_{1}\right)\left(U_{c}\left(\xi\right)-Z_{2}\right)\left(Z_{4}u+Z_{3}\right)\textrm{d}\xi\nonumber \\
 & =c+\epsilon C_{1}\left(C_{2}\sin\left(\frac{2\pi\phi(t)}{L}\right)+C_{3}\cos\left(\frac{2\pi\phi(t)}{L}\right)\right),\end{align}
with

\begin{align}
C_{1} & =-\frac{A\sin\left(\sqrt{2}c\pi\right)}{\left(2c^{3}-c\right)L^{4}\left(\cos\left(2\sqrt{2}c\pi\right)-\cosh\left(\frac{4\sqrt{2}\pi^{2}}{L}\right)\right)},\nonumber \\
C_{2} & =\cosh\left(\frac{2\sqrt{2}\pi^{2}}{L}\right)\sin\left(\sqrt{2}c\pi\right)H_{1}\nonumber \\
 & +2\cos\left(\sqrt{2}c\pi\right)\sinh\left(\frac{2\sqrt{2}\pi^{2}}{L}\right)H_{2},\nonumber \\
C_{3} & =2\cosh\left(\frac{2\sqrt{2}\pi^{2}}{L}\right)\sin\left(\sqrt{2}c\pi\right)H_{2}\nonumber \\
 & -\cos\left(\sqrt{2}c\pi\right)\sinh\left(\frac{2\sqrt{2}\pi^{2}}{L}\right)H_{1},\label{eq:Cs}\end{align}
\begin{multline}
H_{1}=32\pi^{4}Z_{4}+L^{4}\left(2\sqrt{2}\left(\left(2Z_{1}+2Z_{2}-3\right)Z_{4}-2Z_{3}\right)c^{3}\right.\\
\left.\vphantom{2Z_{4}c^{4}+2\sqrt{2}\left(\left(2Z_{1}+2Z_{2}-3\right)Z_{4}-2Z_{3}\right)c^{3}}-c^{2}\left(-12Z_{1}\left(Z_{2}-1\right)+12Z_{2}-11\right)Z_{4}\right.\\
\left.\vphantom{2Z_{4}c^{4}+2\sqrt{2}\left(\left(2Z_{1}+2Z_{2}-3\right)Z_{4}-2Z_{3}\right)c^{3}}+2Z_{4}c^{4}-12c^{2}\left(Z_{1}+Z_{2}-1\right)Z_{3}\right.\\
\left.\vphantom{2Z_{4}c^{4}+2\sqrt{2}\left(\left(2Z_{1}+2Z_{2}-3\right)Z_{4}-2Z_{3}\right)c^{3}}+\sqrt{2}c\left(Z_{1}\left(4-6Z_{2}\right)+4Z_{2}-3\right)Z_{4}\right.\\
\left.\vphantom{2Z_{4}c^{4}+2\sqrt{2}\left(\left(2Z_{1}+2Z_{2}-3\right)Z_{4}-2Z_{3}\right)c^{3}}+\sqrt{2}2c\left(Z_{1}\left(3-6Z_{2}\right)+3Z_{2}-2\right)Z_{3}\right)\\
+L^{2}\left(24\sqrt{2}\pi^{2}\left(2Z_{3}+\left(-2Z_{1}-2Z_{2}+3\right)Z_{4}\right)c\right.\\
\left.\vphantom{24\sqrt{2}\pi^{2}\left(2Z_{3}+\left(-2Z_{1}-2Z_{2}+3\right)Z_{4}\right)c}+4\pi^{2}\left(-12Z_{1}\left(Z_{2}-1\right)+12Z_{2}-11\right)Z_{4}\right.\\
\left.\vphantom{24\sqrt{2}\pi^{2}\left(2Z_{3}+\left(-2Z_{1}-2Z_{2}+3\right)Z_{4}\right)c}+\pi^{2}48\left(Z_{1}+Z_{2}-1\right)Z_{3}-48\pi^{2}Z_{4}c^{2}\right),\label{eq:HsProj1}\end{multline}
\begin{multline}
H_{2}=L8\pi^{3}\left(2\sqrt{2}Z_{3}-\left(4c+\sqrt{2}\left(2Z_{1}+2Z_{2}-3\right)\right)Z_{4}\right)\\
+L^{3}\pi\left(\left(6\sqrt{2}\left(2Z_{1}+2Z_{2}-3\right)c^{2}+\left(24Z_{1}\left(Z_{2}-1\right)-24Z_{2}\right)c\right.\right.\\
\left.\left.+22c+8c^{3}+\sqrt{2}\left(Z_{1}\left(4-6Z_{2}\right)+4Z_{2}-3\right)\right)Z_{4}\right.\\
\left.-2\left(6\sqrt{2}c^{2}+12\left(Z_{1}+Z_{2}-1\right)c\right.\right.\\
\left.\left.+\sqrt{2}\left(-3Z_{2}+Z_{1}\left(6Z_{2}-3\right)+2\right)\right)Z_{3}\right).\label{eq:HsProj2}\end{multline}

\section*{Appendix B}

With the help of the averaging method in 2nd order an ODE for the
position of the front for a sinusoidal variation of the excitation
threshold is derived

\begin{multline}
\Theta_{2}^{A}\left(\phi\right)=H_{1}\sin\left(\frac{2\pi\phi}{L}\right)+H_{2}\cos\left(\frac{2\pi\phi}{L}\right)\\
+H_{3}\sin\left(\frac{4\pi\phi}{L}\right)+H_{4}\cos\left(\frac{4\pi\phi}{L}\right)+H_{5},\end{multline}
\begin{align*}
H_{1} & =\frac{A}{\sqrt{2}D_{1}}\left(\sin\left(2\sqrt{2}c\pi\right)\sinh\left(\frac{2\sqrt{2}\pi^{2}}{L}\right)J_{1}\right.\\
 & \left.\vphantom{\sin\left(2\sqrt{2}c\pi\right)\sinh\left(\frac{2\sqrt{2}\pi^{2}}{L}\right)J_{1}}-4\cosh\left(\frac{2\sqrt{2}\pi^{2}}{L}\right)\sin^{2}\left(\sqrt{2}c\pi\right)J_{3}\right)\\
H_{2} & =\frac{\sqrt{2}A}{D_{1}}\left(\cosh\left(\frac{2\sqrt{2}\pi^{2}}{L}\right)J_{1}\sin^{2}\left(\sqrt{2}c\pi\right)\right.\\
 & \left.\vphantom{\cosh\left(\frac{2\sqrt{2}\pi^{2}}{L}\right)J_{1}\sin^{2}\left(\sqrt{2}c\pi\right)}+\sin\left(2\sqrt{2}c\pi\right)\sinh\left(\frac{2\sqrt{2}\pi^{2}}{L}\right)J_{3}\right)\end{align*}
\begin{align}
H_{3} & =-\frac{A^{2}}{2D_{2}}\left(4\cosh\left(\frac{4\sqrt{2}\pi^{2}}{L}\right)J_{4}\sin^{2}\left(\sqrt{2}c\pi\right)\right.\nonumber \\
 & \left.\vphantom{4\cosh\left(\frac{4\sqrt{2}\pi^{2}}{L}\right)J_{4}\sin^{2}\left(\sqrt{2}c\pi\right)}+\sin\left(2\sqrt{2}c\pi\right)\sinh\left(\frac{4\sqrt{2}\pi^{2}}{L}\right)J_{2}\right)\nonumber \\
H_{4} & =-\frac{A^{2}}{D_{2}}\left(\cosh\left(\frac{4\sqrt{2}\pi^{2}}{L}\right)\sin^{2}\left(\sqrt{2}c\pi\right)J_{2}\right.\nonumber \\
 & \left.\vphantom{\cosh\left(\frac{4\sqrt{2}\pi^{2}}{L}\right)\sin^{2}\left(\sqrt{2}c\pi\right)J_{2}}-\sin\left(2\sqrt{2}c\pi\right)\sinh\left(\frac{4\sqrt{2}\pi^{2}}{L}\right)J_{4}\right)\nonumber \\
H_{5} & =\frac{A^{2}c}{8\pi^{2}}\label{eq:Hsu2}\end{align}

\begin{align}
J_{1} & =c^{2}\left(1-2c^{2}\right)L^{4}+4\left(12c^{2}-1\right)\pi^{2}L^{2}-32\pi^{4},\nonumber \\
J_{2} & =c^{2}\left(2c^{2}-1\right)L^{4}+16\left(1-12c^{2}\right)\pi^{2}L^{2}+512\pi^{4},\nonumber \\
J_{3} & =2c\pi L\left(\left(1-4c^{2}\right)L^{2}+16\pi^{2}\right),\nonumber \\
J_{4} & =4c\pi L\left(\left(1-4c^{2}\right)L^{2}+64\pi^{2}\right),\\
D_{1} & =\left(2c^{3}-c\right)\pi L^{5}\left(\cos\left(2\sqrt{2}c\pi\right)-\cosh\left(\frac{4\sqrt{2}\pi^{2}}{L}\right)\right),\nonumber \\
D_{2} & =4\left(2c^{3}-c\right)\pi^{2}L^{4}\left(\cos\left(2\sqrt{2}c\pi\right)-\cosh\left(\frac{8\sqrt{2}\pi^{2}}{L}\right)\right).\label{eq:Ds}\end{align}

\section*{Appendix C}

For the sinusoidal variation of the fixed point parameter $u_{3},$
2nd order averaging gives an ODE for the position of the front

\begin{multline}
\Theta_{2}^{A}\left(\phi\right)=H_{1}\sin\left(\frac{2\pi\phi}{L}\right)+H_{2}\cos\left(\frac{2\pi\phi}{L}\right)\\
+H_{3}\sin\left(\frac{4\pi\phi}{L}\right)+H_{4}\cos\left(\frac{4\pi\phi}{L}\right)+H_{5},\end{multline}
with\begin{align*}
H_{1} & =-\frac{A}{D_{1}}\sin\left(\sqrt{2}c\pi\right)\left(\cos\left(\sqrt{2}c\pi\right)\sinh\left(\frac{2\sqrt{2}\pi^{2}}{L}\right)J_{1}\right.\\
 & \left.\vphantom{\cos\left(\sqrt{2}c\pi\right)\sinh\left(\frac{2\sqrt{2}\pi^{2}}{L}\right)J_{1}}+\cosh\left(\frac{2\sqrt{2}\pi^{2}}{L}\right)\sin\left(\sqrt{2}c\pi\right)J_{2}\right),\\
H_{2} & =-\frac{A}{D_{1}}\sin\left(\sqrt{2}c\pi\right)\left(\cosh\left(\frac{2\sqrt{2}\pi^{2}}{L}\right)\sin\left(\sqrt{2}c\pi\right)J_{1}\right.\\
 & \left.\vphantom{\cosh\left(\frac{2\sqrt{2}\pi^{2}}{L}\right)\sin\left(\sqrt{2}c\pi\right)J_{1}}-\cos\left(\sqrt{2}c\pi\right)\sinh\left(\frac{2\sqrt{2}\pi^{2}}{L}\right)J_{2}\right),\end{align*}
\begin{align}
H_{3} & =-\frac{A^{2}}{D_{2}}\sin\left(\sqrt{2}c\pi\right)\left(\cosh\left(\frac{4\sqrt{2}\pi^{2}}{L}\right)\sin\left(\sqrt{2}c\pi\right)J_{4}\right.\nonumber \\
 & \left.\vphantom{\cosh\left(\frac{4\sqrt{2}\pi^{2}}{L}\right)\sin\left(\sqrt{2}c\pi\right)J_{4}}-\cos\left(\sqrt{2}c\pi\right)\sinh\left(\frac{4\sqrt{2}\pi^{2}}{L}\right)J_{3}\right),\nonumber \\
H_{4} & =\frac{A^{2}}{D_{2}}\sin\left(\sqrt{2}c\pi\right)\left(\cosh\left(\frac{4\sqrt{2}\pi^{2}}{L}\right)\sin\left(\sqrt{2}c\pi\right)J_{3}\right.\nonumber \\
 & \left.\vphantom{\cosh\left(\frac{4\sqrt{2}\pi^{2}}{L}\right)\sin\left(\sqrt{2}c\pi\right)J_{3}}+\cos\left(\sqrt{2}c\pi\right)\sinh\left(\frac{4\sqrt{2}\pi^{2}}{L}\right)J_{4}\right),\nonumber \\
H_{5} & =\frac{A^{2}c}{8\pi^{2}},\label{eq:Hsu3}\end{align}
\medskip{}
\begin{align}
J_{1} & =c^{2}\left(\sqrt{2}\left(2-3u_{2}\right)+2c\left(\sqrt{2}c+3u_{2}-3\right)\right)L^{4}+32\sqrt{2}\pi^{4}\nonumber \\
 & -4\pi^{2}\left(\sqrt{2}\left(2-3u_{2}\right)+6c\left(2\sqrt{2}c+3u_{2}-3\right)\right)L^{2},\nonumber \\
J_{2} & =4\pi L\left(c\left(c\left(-4\sqrt{2}c-9u_{2}+9\right)+\sqrt{2}\left(3u_{2}-2\right)\right)L^{2}\right.\nonumber \\
 & \left.\vphantom{c\left(c\left(-4\sqrt{2}c-9u_{2}+9\right)+\sqrt{2}\left(3u_{2}-2\right)\right)L^{2}}+4\pi^{2}\left(4\sqrt{2}c+3u_{2}-3\right)\right),\nonumber \\
J_{3} & =c\left(-2c^{3}-6\sqrt{2}\left(u_{2}-1\right)c^{2}+\left(-6\left(u_{2}-3\right)u_{2}-11\right)c\right)L^{4}\nonumber \\
 & +16\pi^{2}\left(12c^{2}+18\sqrt{2}\left(u_{2}-1\right)c+6\left(u_{2}-3\right)u_{2}+11\right)L^{2}\nonumber \\
 & -512\pi^{4}+3\sqrt{2}cL^{4}\left(u_{2}-1\right)^{2},\nonumber \\
J_{4} & =4\pi L\left(\left(-18\sqrt{2}\left(u_{2}-1\right)c^{2}-2\left(6\left(u_{2}-3\right)u_{2}+11\right)c\right)L^{2}\right)\nonumber \\
 & +12\sqrt{2}\pi L^{3}\left(u_{2}-1\right)^{2}-32\pi L^{3}c^{3}\nonumber \\
 & +128\pi^{3}L\left(4c+3\sqrt{2}\left(u_{2}-1\right)\right),\end{align}
and $D_{1},\, D_{2}$ are given as in \eqref{eq:Ds}.

\bibliographystyle{apsrev}
\addcontentsline{toc}{section}{\refname}\bibliography{literature}

\end{document}